\documentclass[twocolumn,superscriptaddress,showpacs,amsmath,amssymb,aps,prl]{revtex4-1}
\pdfoutput=1
\usepackage[pdftex]{graphicx}
\usepackage{amsmath}
\usepackage{braket}

\begin{document}

\title{$\mathrm{Z}_N$ Berry phase and symmetry protected topological phases of SU($N$) antiferromagnetic Heisenberg chain}

\author{Yuichi Motoyama}
\affiliation{Institute for Solid State Physics, University of Tokyo, Kashiwa 277-8581, Japan}
\author{Synge Todo}
\affiliation{Department of Physics, University of Tokyo, Tokyo, 113-0033, Japan}
\affiliation{Institute for Solid State Physics, University of Tokyo, Kashiwa 277-8581, Japan}

\date{\today}

\begin{abstract}
The local Z$_N$ quantized Berry phase for
the SU($N$) antiferromagnetic Heisenberg spin model is formulated.
This quantity, which is a generalization of the local $Z_2$ Berry phase for SU(2) symmetry, has a direct correspondence to the number of
singlet pairs spanning on a particular bond, and is effective as a tool
to characterize and classify the various symmetry protected topological phases
of one-dimensional SU($N$) spin systems.
We extend the path-integral quantum Monte Carlo method
for the Z$_2$ Berry phase in order to calculate the Z$_N$ Berry phase numerically.
We demonstrate our method by calculating the Z$_4$ Berry phase for the bond-alternating SU($4$) antiferromagnetic Heisenberg chain, which is represented by the Young diagram of four columns.
\end{abstract}


\maketitle
The ground state of $S=1$ antiferromagnetic Heisenberg (AFH) chain
is unique and has a finite energy gap,
while that of the classical AFH chain is continuously degenerated and has no energy gap.
The former phase, called Haldane phase~\cite{Haldane1983}, is understood by the valence bond solid (VBS) picture~\cite{AffleckKLT1987}
and characterized by the hidden string order~\cite{NijsR1989}.
The Haldane phase emerges in other one-dimensional systems, such as the spin ladder systems~\cite{MikeskaK2004, HakobyanHR2001},
and has been studied for a long time
by means of several topological order parameters, such as
the string order parameter~\cite{NijsR1989,Oshikawa1992,TodoMYT2001},
the twist order parameter~\cite{NakamuraT2002a},
and the local $\mathrm{Z}_2$ Berry phase~\cite{Hatsugai2006}.

Recently the Haldane phase has been reexamined from a more general viewpoint,
called symmetry protected topological (SPT) phase~\cite{GuW2009}.
In one-dimensional systems with short range interaction,
all topological phases are adiabatically connected to a trivial state (product state) by changing the parameters if one does not impose any symmetries to the Hamiltonian;
some symmetries are required to make topological phases nontrivial and distinguish them from the trivial state.
For example, the Haldane phase of the $S=1$ AFH chain is protected by
any one of the following three symmetries~\cite{PollmannTBO2010}:
time reversal symmetry, bond-centered inversion symmetry,
or the dihedral group of $\pi$ rotation around the $S_x$, $S_y$, and $S_z$ spin axes.
As a promising tool for characterizing such topological phases,
the entanglement spectrum of the ground state has been proposed~\cite{LiH2008}.
This quantity is the spectrum of the reduced density matrix,
which is obtained by tracing out the degrees of freedom
in one part of the whole system from the original density matrix.
In the Haldane state of the one-dimensional AFH $S=1$ chain,
the lowest eigenvalue in the entanglement spectrum is doubly degenerated,
while that of the trivial state is unique~\cite{PollmannTBO2010}.

In the meanwhile, as a candidate that exhibits novel SPT phases,
spin models with generalized SU($N$) symmetries have attracted attentions in years.
In general, the SPT phases of one-dimensional spin systems with some symmetry group $G$
are classified by the second cohomology group $H^2(G,U)$ of $G$ ~\cite{ChenGW2011, ChenGW2011a, DuivenvoordenQ2013}.
It is pointed out that a one-dimensional SU($N$) spin system
can have $N-1$ distinct SPT phases besides the trivial phase
without any additional symmetries~\cite{DuivenvoordenQ2013},
though the SU($2$) $S=1$ spin chain has only one nontrivial topological state, the Haldane state.
If we cut a bond in the SU($N$) spin chain,
SU($N$) edge states emerge at both ends.
The SPT phases of the SU($N$) spin chain can be classified by the number of 
boxes in the Young diagram representing these edge states.
In other words, if we have a means to count the number of boxes directly,
the SPT phases can be identified explicitly.
For example, the degeneracy of the lowest eigenvalues in the entanglement spectrum is the same as the dimension of the representation of the edge state, which helps us to estimate this number.

So far, a number of SU($N$) models have been proposed to study the SPT phases.
Morimoto and his co-workers~\cite{MorimotoUMF2014} constructed
a matrix product state representing an SU($N$) VBS state 
from the viewpoint of SPT
and the corresponding Affleck-Kennedy-Lieb-Tasaki (AKLT) Hamiltonian.
They also calculated the string order parameter~\cite{DuivenvoordenQ2012}
of the SU($3$) AKLT model by the density matrix renormalization group method.
Geraedts and Motrunich~\cite{GeraedtsM2014arXiv} proposed another exactly solvable SU($N$) model that has $N$ distinct SPT phases as the ground state
by generalizing the cluster Ising model~\cite{PachosP2004}.
Quest for SU($N$) SPT states in the real world is also in progress.
Ultracold alkaline-earth atom systems, such as ${}^{87}$Sr, ${}^{171}$Yb, or ${}^{173}$Yb atoms,
in an optical lattice are described by a two-orbital SU($N$) Hubbard model
or an SU($N$) AFH model in the Mott insulating phase~\cite{GorshkovHGXJYZDLR2010}.
Nonne and his co-workers studied the former model analytically and numerically,
and found SU($N$) SPT phase~\cite{NonneMCLT2013}.

In this Letter, we introduce the local $\mathrm{Z}_N$ Berry phase~\cite{HatsugaiM2011} to the SU($N$) AFH model
and show that this quantity can detect the number of SU($N$) singlet pairs on a particular bond.
This number corresponds directly to the number of boxes
in the Young diagram of the edge states,
and thus enables us to distinguish several SPT phases.
We then extend the path-integral quantum Monte Carlo (PIQMC) method
for calculating the local $\mathrm{Z}_2$ Berry phase~\cite{MotoyamaT2013} to the present $\mathrm{Z}_N$ case.
To demonstrate the power of the Berry phase and our PIQMC method,
we perform a simulation of the bond-alternating SU($4$) AFH chain, which is represented by 
a Young diagram of four columns,
and confirm that the local $\mathrm{Z}_4$ Berry phase indeed identifies the nontrivial SPT phases correctly.

The SU($N$) antiferromagnetic spin model can be generally written as
\begin{equation}
  \mathcal{H} = \sum_{\left\langle i,j \right\rangle} 
  \sum_{\alpha, \beta = 1}^N J_{ij} S^\alpha_\beta(i)S^\alpha_\beta(j),
\end{equation}
where $\left\langle i,j\right\rangle$ represents a nearest-neighbor bond connecting
sites $i$ and $j$, and the outer summation runs over all the nearest-neighbor bonds.
$S^\alpha_\beta$ denotes a generator of SU($N$) algebra
satisfying the following commutation relation:
$[ S^\alpha_\beta(i), S^\mu_\nu(j) ] 
= \delta_{i,j}( \delta^\alpha_\nu S^\mu_\beta(i) - \delta^\mu_\beta S^\alpha_\nu(i)).$
In this Letter, we consider bipartite lattices and
adopt the representation of the algebra depicted
as a Young diagram of one row and $M$ columns for one sublattice $A$
and the one depicted as one of $(N-1)$ rows and $M$ columns for the other sublattice $B$~\cite{Affleck1985a, ReadS1990, HaradaKT2003}.
For $M=1$, the model can be written in terms of an $N$-color particle basis as
\begin{equation}
  \mathcal{H} =
  -\sum_{i \in A} \sum_{j \in B}
  \sum_{\alpha_i, \alpha_j, \beta_i, \beta_j}
  \!\!\!\!\!\! J_{ij} \ket{\alpha_i \tilde{\alpha}_j}
  \frac{\delta_{\alpha_i, \alpha_j} \delta_{\beta_i, \beta_j}}{N}
  \bra{\beta_i \tilde{\beta}_j},
\end{equation}
where $\ket{\alpha_i} = c^\dagger_{\alpha,i}\ket{\text{vacuum}}$
and $c^\dagger_{\alpha,i}$ is the fermion creation operator of color $\alpha$ at site $i,$
and $\ket{\tilde{\alpha}_j} = c_{\alpha,j} \prod_{\beta} c_{\beta,j}^\dagger\ket{\text{vacuum}} = \left(-1\right)^\alpha \otimes_{\beta \ne \alpha}\ket{\beta_j}.$
Physically, $\ket{\alpha}$ and $\ket{\tilde{\alpha}}$ mean 
an $\alpha$ particle and hole, respectively~\cite{Affleck1985a}.
When $J_{ij}>0,$ the ground state of the dimer Hamiltonian has a unique ground state
$\ket{\psi} = N^{-1/2}\sum_{\alpha=1}^N \ket{\alpha \tilde{\alpha}}$ with
energy $E=-1$ and $(N^2-1)$-fold degenerated excited states with $E=0$.
We call this ground state as the SU($N$) singlet.

The $\alpha_1 = \alpha_2$ subspace of the Hamiltonian of SU($N$) AFH dimer,
which is represented by an $N \times N$ matrix with all the elements being $-1/N$
and still includes the SU($N$) singlet state, is
the same as the Hamiltonian of a hopping particle on $N$ fully connected sites.
Since the ground state of the latter
is characterized by the $\mathrm{Z}_N$ Berry phase~\cite{HatsugaiM2011},
so is the SU($N$) singlet bond.
In the following, we will introduce the $\mathrm{Z}_N$ Berry phase $\gamma$ for the present SU($N$) AFH model as in Ref.~\cite{HatsugaiM2011}
and show explicitly that the value of the Berry phase of the SU($N$) singlet is given by $\gamma = -2\pi/N \mod 2\pi$.
First, we introduce $(N-1)$ parameters $\theta_\alpha$ with $\alpha = 1,2,\dots,N-1$ and the twist vector $\vec{\phi}$, whose elements are defined as
\begin{equation}
  \phi_\alpha = \begin{cases}
    \sum_{\beta=1}^\alpha \theta_\beta & 1\le \alpha < N \\
    0 & \alpha = N.
  \end{cases}
\end{equation}
By using $\vec{\phi},$ we ``twist'' the dimer Hamiltonian as
\begin{equation}
  \mathcal{H}(\vec\phi) = - \sum_{\alpha, \beta}
  \ket{\alpha \tilde{\alpha}}
  \frac{ e^{-i(\phi_{\alpha} - \phi_{\beta})} }{N}
  \bra{\beta \tilde{\beta}}.
  \label{eq:SUN_twisted}
\end{equation}
The corresponding ground state is also twisted as 
$\ket{\psi(\vec\phi)} = N^{-1/2} \sum_\alpha \exp(-i\phi_\alpha) \ket{\alpha\tilde{\alpha}}$.
Next, we parameterize $\{\theta_\alpha\}$ by introducing a ``time'' parameter $t$ as follows:
\begin{equation}
  \theta_1 = t
\end{equation}
and 
\begin{equation}
  \theta_\alpha = \begin{cases}
    t & 0 \le t < 2\pi/N \\
  \frac{2\pi-t}{N-1} & 2\pi/N \le t < 2\pi 
\end{cases}
\end{equation}
for $1<\alpha<N.$
By using this parameterization, we define the Berry curvature $A_\alpha(\vec\phi)$ as
\begin{equation}
  A_\alpha(\vec\phi) = \Braket{\psi(\vec\phi)|\frac{\partial}{\partial \phi_\alpha}|\psi(\vec\phi)},
\end{equation}
and the Berry phase $\gamma$ as
\begin{equation}
  \gamma = i\oint \mathrm{d}\vec\phi \cdot \vec{A}(\vec\phi)
  = i\int_0^{2\pi} \!\!\!\! \mathrm{d}t \sum_\alpha \frac{\mathrm{d}\phi_\alpha}{\mathrm{d}t} A_\alpha(\vec\phi(t)).
\end{equation}
It should be noted that $\phi_\alpha(2\pi) = 0 \mod 2\pi$ and
$\mathcal{H}(\vec\phi(2\pi)) = \mathcal{H}(\vec\phi(0))$ make the parameter path $\vec\phi(t)$ a closed loop
and accordingly the Berry phase becomes gauge invariant.
Finally, the Berry curvature and the Berry phase of the SU($N$) dimer can be evaluated explicitly as
\begin{equation}
  A_\alpha(\vec\phi) = \frac{1}{N} \sum_{\beta,\beta'}
  \Braket{\beta \tilde{\beta}|e^{i\phi_\beta}\frac{\partial}{\partial \phi_\alpha}e^{-i\phi_{\beta'}}|\beta'\tilde{\beta'}}
  = -\frac{i}{N}
\end{equation}
and
\begin{equation}
  \begin{split}
    \gamma &= i\int_0^{2\pi} \!\!\!\! \mathrm{d}t \sum_{\alpha=1}^N \frac{\mathrm{d}\phi_\alpha}{\mathrm{d}t} A_\alpha(\vec\phi)
           = \frac{1}{N}\int_0^{2\pi} \!\!\!\! \mathrm{d}t \sum_{\alpha=1}^{N-1} \sum_{\beta=1}^{\alpha} \frac{\mathrm{d}\theta_\beta}{\mathrm{d}t} \\
           &= 2\pi\frac{N-1}{N},
  \end{split}
\end{equation}
respectively, where we used relations $\phi_N = 0$ and $\theta_\beta(2\pi) - \theta_\beta(0) = 2\pi\delta_{\beta,1}$.

For $M>1$, a spin can be regarded as a symmetric summation
of $M$ one-column SU($N$) subspins as an SU($2$) spin with $S>1/2$
can be decomposed into $2S$ subspins with $S=1/2$.
Thus, the ground state of $M$-column SU($N$) dimer is
the symmetric superposition of $M!$ product states of $M$ singlet pairs as
\begin{equation}
\begin{split}
  \ket{\Psi(\vec\phi)}
  &= \frac{1}{\sqrt{M!}} \sum_P \ket{\Psi_P(\vec\phi)} \\
  &= \frac{1}{\sqrt{M!}} \sum_P 
  \bigotimes_{f=1}^M \frac{1}{\sqrt{N}} \sum_{\alpha=1}^N e^{-i\phi_\alpha} \ket{\alpha_f \tilde{\alpha}_{P(f)}},
\end{split}
\end{equation}
where $f$ is the index of subspins and $P$ denotes a permutation of $M$ distinct elements.
A short explicit calculation gives that the cross term
in the Berry connection,
$\braket{\Psi_P|\partial_{\phi_\alpha}|\Psi_{P'}},$
becomes $-iM/N \braket{\Psi_P|\Psi_{P'}}$
\footnote{
  The differentiation erases a summation as $\partial_{\phi_\alpha}\sum_\beta e^{-i\phi_\beta} = -ie^{-i\phi_\alpha},$ and produces the denominator, $N$.
  The numerator, $M$, comes from the Leibniz's product rule.
}.
The Berry connection and the Berry phase are obtained in the same way as the $M=1$ case as
$ A_\alpha(\vec\phi) =
\braket{\Psi(\vec\phi)|\partial_\alpha|\Psi(\vec\phi)}
= -iM/N $
and $\gamma = 2\pi M (N-1)/N,$ respectively.

The value of the local Berry phase on bond $b$ thus
gives the number of singlet pairs on the bond
as $n(b) = \gamma(b) N / (2\pi(N-1))$.
If we cut bond $b$, $n(b)$ singlets are removed and
$2n(b)$ subspins are left as the edge state.
The edge state on the site belonging to sublattice $A$
is depicted as the Young diagram of $1 \times n(b)$ boxes.
For one-dimensional spin systems,
the number of boxes characterizes SPT phases~\cite{DuivenvoordenQ2013},
and so does the local Z$_N$ Berry phase.

In the following, we will describe how to calculate
the Berry connection and the Berry phase by means of the PIQMC method.
The Berry connection, $A(t) = \braket{\psi(t)|\partial_t|\psi(t)}$,
is the coefficient of the first order term of 
the series expansion of the inner product between
two normalized ground states:
\begin{equation}
\braket{\psi(t)|\psi(t+\delta t)}
= 1 + \delta t A(t) + O(\delta t^2).
\label{eq:inner_product_A}
\end{equation}
The ground state $\ket{\psi(t)}$ of the Hamiltonian $\mathcal{H}(t)$ can be obtained by the projection method as
$\ket{\Psi(t)} = \lim_{\beta \to \infty} N(t, \beta) \exp\left(-\beta \mathcal{H}(t)/2\right) \ket{\phi},$
where $\beta$ is the projection parameter, $N(t,\beta) \in \mathbb{R}$ a normalization factor,
and $\ket{\phi}$ some initial state nonorthogonal to the ground state.
We will omit the ``$\lim$'' symbol for simplicity from now.
The inner product of two ground states (\ref{eq:inner_product_A}) is also represented by the projection as
$\braket{\psi(t)|\psi(t+\delta t)} = N(t)N(t+\delta t) \braket{\phi|\exp\left(-\beta \mathcal{H}(t)/2\right)\exp\left(-\beta \mathcal{H}(t+\delta t)/2\right)|\phi}.$

Next, as one expands $\mathrm{Tr}\exp(-\beta\mathcal{H})$ in the ordinary PIQMC,
we expand $\braket{\phi|\exp(\dots)|\phi}$ by path integrals as
\begin{equation}
  \braket{\psi(t)|\psi(t+\delta t)} = N(t)N(t+\delta t) \sum_c W(c; t, t+\delta t),
  \label{eq:inner_product_B}
\end{equation}
where $W(c; t, t+\delta t)$ is the weight of a worldline configuration, $c$.
By using the identities, $N(t)^2\sum_c W(c;t,t) = \braket{\psi(t)|\psi(t)} = 1$ and
$N(t)\frac{\mathrm{d}}{\mathrm{d}t}N(t) = \frac{\mathrm{d}}{\mathrm{d}t}N(t)^2/2,$
Eq.\,(\ref{eq:inner_product_B}) can be expanded with respect to $\delta t$
up to the first order as
\begin{equation}
\begin{split}
  &\Braket{\psi(t)|\psi(t+\delta t)} \\
  & \qquad \simeq 1 + \frac{\delta t}{2}\frac{\sum_c(\partial_{t_\text{L}} - \partial_{t_\text{U}})W(c;t_\text{U}, t_\text{L})}{\sum_c W(c;t,t)} \Big|_{t_\text{L}=t_\text{U} = t} \\
  & \qquad = 1 + \delta t \frac{\Braket{\partial_{t_\text{L}}} - \Braket{\partial_{t_\text{U}}}}{2},
\label{eq:inner_product_C}
\end{split}
\end{equation}
where $\braket{O} = \sum_c O(c) W(c,t,t) / \sum_c W(c,t,t).$
Combining Eq.\,(\ref{eq:inner_product_C}) with Eq.\,(\ref{eq:inner_product_A})
we obtain a generic form for the Berry connection,
\begin{equation}
  A(t) = \frac{1}{2}\bigl( \Braket{\partial_{t_\text{L}}}
  - \Braket{\partial_{t_\text{U}}}\bigr).
  \label{eq:BC_evaluator}
\end{equation}
Note that since we do not use any knowledge of any specific models so far,
this formula is general and valid for any models.

To advance the procedure further,
we now need an explicit representation for $\braket{\partial}$
based on the weight function $W(c)$ of the model under consideration.
The expectation of derivative with respective to $t$ can be evaluated in the same way as the energy evaluator, $\braket{\mathcal{H}} = -\braket{\partial_\beta} = -\braket{m}/\beta$,
where $m$ is the number of operators~\cite{Todo2013a}.
For the present SU($N$) AFH model~(\ref{eq:SUN_twisted}),
the evaluator of the Berry connection is eventually written as
\begin{equation}
  A_\alpha(c; \vec\phi) = -\frac{i}{2}\sum_{\beta=1}^N
    [( N_{\alpha\beta}^{\text{L}} - N_{\beta\alpha}^{\text{L}} )
    - ( N_{\alpha\beta}^{\text{U}} - N_{\beta\alpha}^{\text{U}} )],
\end{equation}
where $N_{\alpha\beta}^{\text{L(U)}}$ is the number of the operators,
$\ket{\alpha\tilde{\alpha}}\bra{\beta\tilde{\beta}}$, on the twisted bond
at the imaginary time $\tau \in [0,\beta/2] \  (\tau \in [\beta/2,\beta])$.
\begin{figure}[tb]
\begin{center}
\includegraphics[width=\linewidth]{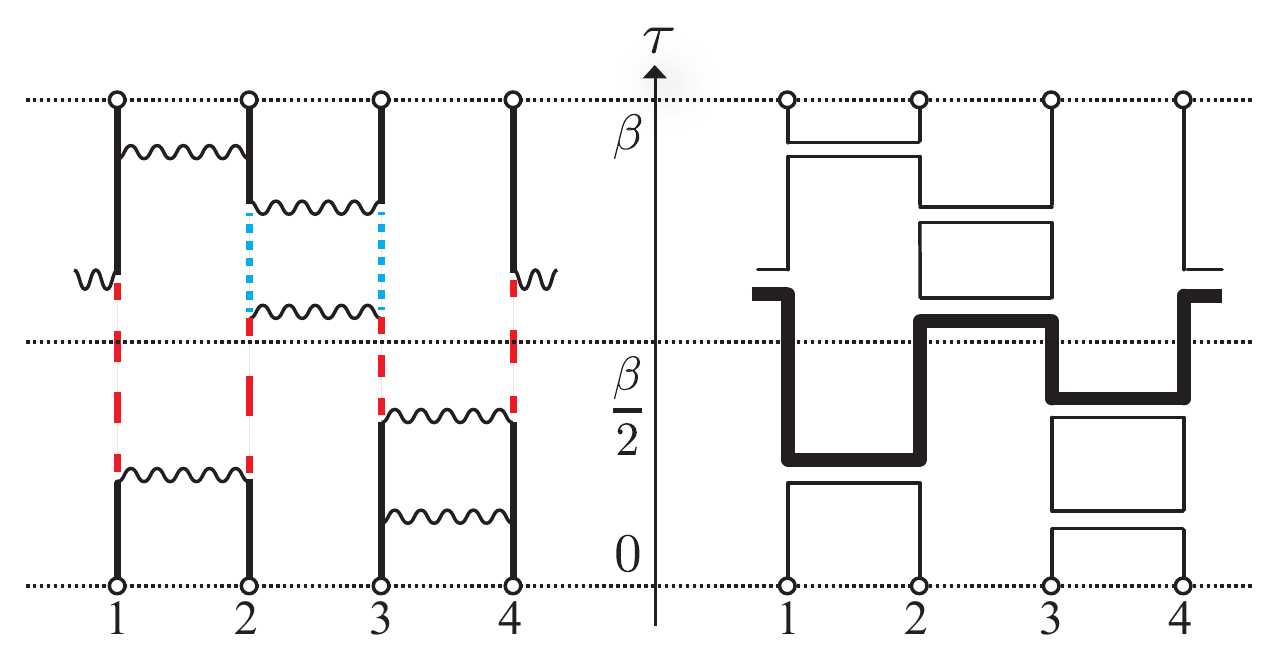}
\end{center}
\caption{
  (Color online) Examples of worldline configuration (left panel)
  and loop configuration (right panel) of the SU($N$) AFH model.
  Color of worldlines denotes spin state;
  black line ($N$), red dashed line ($1$), and blue dotted line ($2$).
}
\label{fig:SUN_worldline}
\end{figure}
The left panel of Fig.~\ref{fig:SUN_worldline} shows
an example of worldline configuration of four site system.
Black lines, red dashed lines, and blue dotted lines denote
spin state $N, 1$, and $2$, respectively.
When we twist bond $\braket{1,2},$ $N_{1,N}^\text{L} = N_{N,N}^{\text{U}} = 1$ and 
the other $N_{\alpha\beta}^\text{L}$'s and $N_{\alpha\beta}^\text{U}$'s
are zero in this configuration.
Note that the twist perturbation makes the weights complex-valued,
so we should use the reweighting method to perform Monte Carlo simulation
as $\braket{A} = \braket{AS}_0/\braket{S}_0$,
where $\braket{\cdot}_0$ is the expectation value
under the absolute weight system with $W_0 = |W|$,
and $S = W/|W|$ is the phase function written as
$S(c; \vec\phi) = \prod_{\alpha, \beta}\exp[ -i(\phi_\alpha - \phi_\beta) N_{\alpha\beta}]$
for the present model.
Berry connections for all parameters can be calculated
from one Monte Carlo simulation,
and the Berry phase can be obtained by integrating these numerically.

For updating the worldline configuration, we employ the loop cluster algorithm~\cite{BeardW1996,TodoK2001,Todo2013a}. In the loop cluster algorithm, one can represent the Berry connection as a function of loop configuration (improved estimator) instead of the worldline configuration~\cite{MotoyamaT2013}.
It should be noted that in the present projector Monte Carlo, the states at $\tau = 0$ and $\beta$ are fixed
to some reference state, $\ket{\phi}$,
which break edge loops and prevent these from flipping.
We choose $\ket{\phi}$ as the ``N\'{e}el'' state, $\otimes_i \ket{N}_i$,
as this choice makes the improved estimators simple because of $\phi_N = 0$ for any $t$.
In the improved estimator, the Berry connection and the phase of the weight
are decomposed into the contributions from each closed loop as
$ 2iA_\alpha = \sum_\ell n_-(\ell) \delta_{\alpha(\ell), \alpha}$
and $S = \prod_\ell \exp{i \phi_{\alpha(\ell)} n_+(\ell)}$, respectively,
where $\alpha(\ell)$ is the state (color) of loop $\ell$ and
$ n_\pm = \left(N_{\text{L}, \uparrow} - N_{\text{L}, \downarrow}\right)
\pm \left( N_{\text{U}, \uparrow} - N_{\text{U}, \downarrow}\right), $
where these $N$'s are the numbers of the jumps
above ($\uparrow$) and below ($\downarrow$) the operator 
at $\tau < \beta/2$ (L) and $\tau>\beta/2$ (U) on the twisted bond.
By tracing out the spin state, we obtain the expression of the improved estimator for the Berry phase as
\begin{widetext}
\begin{equation}
  \gamma 
  = i\int_0^{2\pi}\!\!\!\mathrm{d}t \, \frac{\mathrm{d}\vec{\phi}(t)}{\mathrm{d}t}\cdot \vec{A}(\vec{\phi}(t))
  =
  \int_0^{2\pi}\frac{\mathrm{d}t}{2}
  \Braket{
    \sum_{\ell}
    \sum_\alpha
    \frac{\mathrm{d}\phi_\alpha(t)}{\mathrm{d}t}
    \frac{n_-(\ell)e^{-in_+(\ell)\phi_\alpha}}{N}
    \prod_{\ell'\ne\ell}
    \sum_\alpha
    \frac{e^{-in_+(\ell')\phi_\alpha}}{N}
  }_0
  \Bigg/
  \Braket{
    \prod_{\ell}
    \sum_\alpha
    \frac{e^{-in_+(\ell)\phi_\alpha}}{N}
  }_0,
  \label{eq:BP_improved}
\end{equation}
\end{widetext}
where the loop indices $\ell$ and $\ell'$ run over all closed loops.
For $N=2,$ this estimator is reduced to the one derived in the past work~\cite{MotoyamaT2013}.
A typical loop configuration for the SU($N$) AFH model is shown in the right panel of
Fig.~\ref{fig:SUN_worldline}.
If we twist bond $\braket{1,2}$, $n_\pm$ of the thick loop are both $1$,
and those of the other loops are all $0$ in this case.

We can extend the above procedure straightforwardly to higher-order terms in Eq.\,(\ref{eq:inner_product_A}), which yields estimators for other useful quantities.
From the second order, for example, the expressions for the susceptibility of fidelity~\cite{QuanSLZS2006}
$\chi_F = \lim_{\delta t \to 0}\left|\braket{\psi(t)|\psi(t+\delta t}\right|^2/(\delta t)^2$
and the Berry curvature
$\Omega_{\alpha\beta} = \partial_\alpha A_\beta - \partial_\beta A_\alpha$
are derived as
$ \chi_F = \braket{\partial_\text{U}\partial_\text{L}}
       - \braket{\partial_\text{U}}\braket{\partial_\text{L}} $
and 
$ \Omega_{\alpha\beta} =
(
\braket{\partial_{\alpha,\text{U}}\partial_{\beta,\text{L}}}
\braket{\partial_{\alpha,\text{U}}}\braket{\partial_{\beta,\text{L}}}
) - ( \alpha \leftrightarrow \beta)$, respectively.
We note that more specialized QMC estimators
have been proposed for the susceptibility of the fidelity~\cite{SchwandtAC2009}
and for the Berry curvature~\cite{Kolodrubetz2014}, which are more accurate for the specific models.

In order to demonstrate the ability of the local Z$_N$ Berry phase, we applied our PIQMC method to the 4-column SU(4) AFH model ($N=4$ and $M=4$) on a bond alternating chain~\cite{Affleck1985a}.
The Hamiltonian is written as
\begin{equation}
  \mathcal{H}(\delta) = \sum_{i=1}^L \left(1 - (-1)^i \delta \right) \mathcal{H}_{i,i+1},
\end{equation}
where $\mathcal{H}_{i,i+1}$ is the two-body Hamiltonian on a bond connecting
$i$-th and $(i+1)$-th sites, $L$ the length of chain,
and the periodic boundary condition, $S^\alpha_\beta(i+L) = S^\alpha_\beta(i)$, is imposed.
In the $\delta = 1$ limit, the ground state is obviously
the product state of decoupled $L/2$ dimers so that
the number of singlet pairs on a $1+\delta$ bond, $n(+)$, is $4$
and one on a $1-\delta$ bond, $n(-)$, is $0.$
In the $\delta = -1$ limit, on the other hand,
$n(+)$ and $n(-)$ are $0$ and $4,$ respectively.
As $\delta$ is increased from $-1$ to $1$,
it is expected that quantum phase transitions between different topological phases occur successively.
We calculated the Berry phase $\gamma$ by numerically integration
of the Berry connections $A(t)$ evaluated by the present PIQMC method at twist parameters $t = 2\pi k/32$ with $k = 0,1,\dots,31$.
For the chain lattice, $n_+(\ell)$ in Eq.\,(\ref{eq:BP_improved}) is almost always zero
because $n_+(\ell)$ is nothing but the winding number of the loop.
As a result, the phase factor of the weight, $S$, stays almost always $1$,
and allows us to perform a precise simulation without introducing any additional technique for solving
the sign problem, such as the meron cluster algorithm~\cite{ChandrasekharanW1999, MotoyamaT2013}.
We also calculated the ``staggered'' susceptibility, $\chi$, the response function against the field
that favors the N\'{e}el state, i.e., $\otimes_{i\in A}\otimes_{j\in B}\ket{\alpha}_i \ket{\tilde{\alpha}}_j$, by using the standard loop algorithm.
The projection parameter for calculating the Berry phase and the inverse temperature for the susceptibility are both chosen as $\beta = 2L$.

\begin{figure}[tb]
  \centering
  \includegraphics[width=\linewidth]{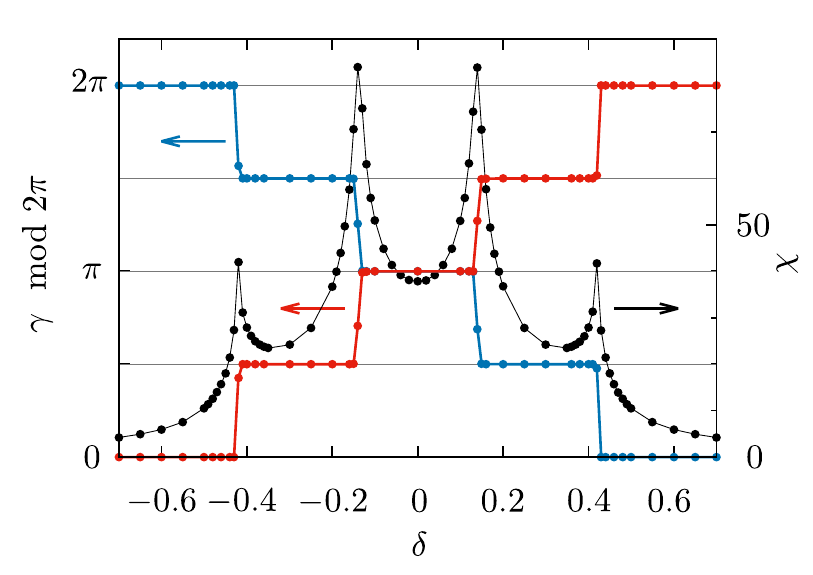}
  \caption{
    (Color online)
    $\delta$-dependence of the Berry phase (left axis) and the staggered susceptibility (right axis)
    of the bond-alternating 4-column SU(4) AFH chain with $L=64$.
    Red symbols forming climbing stairs (blue forming down stairs) denote the Berry phase $\gamma$ on a $1-\delta$ ($1+\delta$) bond.
    Black symbols denote the staggered susceptibility $\chi$.
    Both the projection parameter and the inverse temperature are $\beta=2L$.
    The error bar is smaller than the symbol size.
  }
  \label{fig:bp_ssus}
\end{figure}

Figure~\ref{fig:bp_ssus} shows the simulation results for the $L=64$ chain. It is clear that the staggered susceptibility tends to diverge at $\delta \simeq -0.42$, $-0.14$, $0.14$, and 0.42, suggesting quantum phase transitions between different phases. Although such successive phase transitions have already been observed in the SU(2) spin system with $M=4$, i.e., $S=2$ bond-alternating Heisenberg chain~\cite{NakamuraT2002a}, $\delta$-dependence of the Berry phase in Fig.~\ref{fig:bp_ssus} elucidates that the nature of the intermediated phases are completely different from that in the SU(2) case.  On the $1-\delta$ bond (red symbols in Fig.~\ref{fig:bp_ssus}), the Berry phase shows a step-like behavior and the height of each step is $-3\pi/2$ (or $\pi/2$ mod $2\pi$), which corresponds to the decrease of the number of singlet pairs on the bond, while on a $1+\delta$ bond (blue symbols), the Berry phase decreases by $\pi/2$ (or increases by $3\pi/2$) at each critical point.
Since the Berry phase is quantized by symmetry as long as the energy gap is opened,
two gapped states with different Berry phase are topologically different~\cite{Hatsugai2006}.
Thus, we conclude that the present model exhibits four different topological phases with $\gamma = 0$, $\pi/2$, $\pi$, and $3\pi/2$, respectively. This is in sharp contract with the $S=2$ SU(2) case, where only one nontrivial topological phase exists besides the trivial one.

In summary, we defined the local Z$_N$ Berry phase 
for the SU($N$) antiferromagnetic Heisenberg model,
which can be directly related to the number of the singlet pairs on a bond.
We also extended the path-integral quantum Monte Carlo method
for local Z$_2$ Berry phase of SU($2$) Heisenberg model
to general models,
and derived the evaluator of the Z$_N$ Berry phase
for the SU($N$) antiferromagnetic Heisenberg model.
For demonstration, we calculated the Berry phase
of the SU($4$) antiferromagnetic Heisenberg model for the representation
depicted as Young diagrams of $4$ columns on bond alternating chains. Note that in the $M=4$ SU(4) case the degrees of freedom on one spin is 35, which corresponds to 5.13 $S=1/2$ spins in the SU(2) case. The rapid increase of the size of the Hilbert space makes it impossible to perform simulations based on wave functions, such as the exact diagonalization and the DMRG method.
Although the classical quantities such as the susceptibility can not identify different topological phases, the Berry phase successfully reveals that the present model indeed has $N$ distinct SPT phases.

The simulation code used in the present study has been developed based on the ALPS library~\cite{ALPS2011,ALPSweb}.
The authors thank the Supercomputer Center, the Institute for Solid State Physics,
the University of Tokyo for the facilities and the use of the SGI Altix ICE 8400EX.
The authors acknowledge support by the Computational Materials Science initiative (CMSI) in the HPCI Strategic Programs for Innovative Research (SPIRE) from MEXT, Japan.
S.T. acknowledges the support by KAKENHI (No.\,23540438, 26400384) from JSPS.

\bibliography{bp}

\begin{thebibliography}{36}%
\makeatletter
\providecommand \@ifxundefined [1]{%
 \@ifx{#1\undefined}
}%
\providecommand \@ifnum [1]{%
 \ifnum #1\expandafter \@firstoftwo
 \else \expandafter \@secondoftwo
 \fi
}%
\providecommand \@ifx [1]{%
 \ifx #1\expandafter \@firstoftwo
 \else \expandafter \@secondoftwo
 \fi
}%
\providecommand \natexlab [1]{#1}%
\providecommand \enquote  [1]{``#1''}%
\providecommand \bibnamefont  [1]{#1}%
\providecommand \bibfnamefont [1]{#1}%
\providecommand \citenamefont [1]{#1}%
\providecommand \href@noop [0]{\@secondoftwo}%
\providecommand \href [0]{\begingroup \@sanitize@url \@href}%
\providecommand \@href[1]{\@@startlink{#1}\@@href}%
\providecommand \@@href[1]{\endgroup#1\@@endlink}%
\providecommand \@sanitize@url [0]{\catcode `\\12\catcode `\$12\catcode
  `\&12\catcode `\#12\catcode `\^12\catcode `\_12\catcode `\%12\relax}%
\providecommand \@@startlink[1]{}%
\providecommand \@@endlink[0]{}%
\providecommand \url  [0]{\begingroup\@sanitize@url \@url }%
\providecommand \@url [1]{\endgroup\@href {#1}{\urlprefix }}%
\providecommand \urlprefix  [0]{URL }%
\providecommand \Eprint [0]{\href }%
\providecommand \doibase [0]{http://dx.doi.org/}%
\providecommand \selectlanguage [0]{\@gobble}%
\providecommand \bibinfo  [0]{\@secondoftwo}%
\providecommand \bibfield  [0]{\@secondoftwo}%
\providecommand \translation [1]{[#1]}%
\providecommand \BibitemOpen [0]{}%
\providecommand \bibitemStop [0]{}%
\providecommand \bibitemNoStop [0]{.\EOS\space}%
\providecommand \EOS [0]{\spacefactor3000\relax}%
\providecommand \BibitemShut  [1]{\csname bibitem#1\endcsname}%
\let\auto@bib@innerbib\@empty
\bibitem [{\citenamefont {Haldane}(1983)}]{Haldane1983}%
  \BibitemOpen
  \bibfield  {author} {\bibinfo {author} {\bibfnamefont {F.~D.~M.}\
  \bibnamefont {Haldane}},\ }\href@noop {} {\bibfield  {journal} {\bibinfo
  {journal} {Phys. Lett. A}\ }\textbf {\bibinfo {volume} {93}},\ \bibinfo
  {pages} {464} (\bibinfo {year} {1983})}\BibitemShut {NoStop}%
\bibitem [{\citenamefont {Affleck}\ \emph {et~al.}(1987)\citenamefont
  {Affleck}, \citenamefont {Kennedy}, \citenamefont {Lieb},\ and\ \citenamefont
  {Tasaki}}]{AffleckKLT1987}%
  \BibitemOpen
  \bibfield  {author} {\bibinfo {author} {\bibfnamefont {I.}~\bibnamefont
  {Affleck}}, \bibinfo {author} {\bibfnamefont {T.}~\bibnamefont {Kennedy}},
  \bibinfo {author} {\bibfnamefont {E.~H.}\ \bibnamefont {Lieb}}, \ and\
  \bibinfo {author} {\bibfnamefont {H.}~\bibnamefont {Tasaki}},\ }\href@noop {}
  {\bibfield  {journal} {\bibinfo  {journal} {Phys. Rev. Lett.}\ }\textbf
  {\bibinfo {volume} {59}},\ \bibinfo {pages} {799} (\bibinfo {year}
  {1987})}\BibitemShut {NoStop}%
\bibitem [{\citenamefont {den Nijs}\ and\ \citenamefont
  {Rommelse}(1989)}]{NijsR1989}%
  \BibitemOpen
  \bibfield  {author} {\bibinfo {author} {\bibfnamefont {M.}~\bibnamefont {den
  Nijs}}\ and\ \bibinfo {author} {\bibfnamefont {K.}~\bibnamefont {Rommelse}},\
  }\href@noop {} {\bibfield  {journal} {\bibinfo  {journal} {Phys. Rev. B}\
  }\textbf {\bibinfo {volume} {40}},\ \bibinfo {pages} {4709} (\bibinfo {year}
  {1989})}\BibitemShut {NoStop}%
\bibitem [{\citenamefont {Mikeska}\ and\ \citenamefont
  {Kolezhuk}(2004)}]{MikeskaK2004}%
  \BibitemOpen
  \bibfield  {author} {\bibinfo {author} {\bibfnamefont {H.-J.}\ \bibnamefont
  {Mikeska}}\ and\ \bibinfo {author} {\bibfnamefont {A.~K.}\ \bibnamefont
  {Kolezhuk}},\ }in\ \href@noop {} {\emph {\bibinfo {booktitle} {Quantum
  Magnetism}}},\ \bibinfo {editor} {edited by\ \bibinfo {editor} {\bibfnamefont
  {U.}~\bibnamefont {Schollw\"{o}ck}}, \bibinfo {editor} {\bibfnamefont
  {J.}~\bibnamefont {Richter}}, \bibinfo {editor} {\bibfnamefont {D.~J.}\
  \bibnamefont {Farnell}}, \ and\ \bibinfo {editor} {\bibfnamefont {R.~F.}\
  \bibnamefont {Bishop}}}\ (\bibinfo  {publisher} {Springer-Verlag},\ \bibinfo
  {address} {Berlin},\ \bibinfo {year} {2004})\ pp.\ \bibinfo {pages}
  {1--83}\BibitemShut {NoStop}%
\bibitem [{\citenamefont {Hakobyan}\ \emph {et~al.}(2001)\citenamefont
  {Hakobyan}, \citenamefont {Hetherington},\ and\ \citenamefont
  {Roger}}]{HakobyanHR2001}%
  \BibitemOpen
  \bibfield  {author} {\bibinfo {author} {\bibfnamefont {T.}~\bibnamefont
  {Hakobyan}}, \bibinfo {author} {\bibfnamefont {J.~H.}\ \bibnamefont
  {Hetherington}}, \ and\ \bibinfo {author} {\bibfnamefont {M.}~\bibnamefont
  {Roger}},\ }\href@noop {} {\bibfield  {journal} {\bibinfo  {journal} {Phys.
  Rev. B}\ }\textbf {\bibinfo {volume} {63}},\ \bibinfo {pages} {144433}
  (\bibinfo {year} {2001})}\BibitemShut {NoStop}%
\bibitem [{\citenamefont {Oshikawa}(1992)}]{Oshikawa1992}%
  \BibitemOpen
  \bibfield  {author} {\bibinfo {author} {\bibfnamefont {M.}~\bibnamefont
  {Oshikawa}},\ }\href@noop {} {\bibfield  {journal} {\bibinfo  {journal} {J.
  Phys. Condens. Matter}\ }\textbf {\bibinfo {volume} {4}},\ \bibinfo {pages}
  {7469} (\bibinfo {year} {1992})}\BibitemShut {NoStop}%
\bibitem [{\citenamefont {Todo}\ \emph {et~al.}(2001)\citenamefont {Todo},
  \citenamefont {Matsumoto}, \citenamefont {Yasuda},\ and\ \citenamefont
  {Takayama}}]{TodoMYT2001}%
  \BibitemOpen
  \bibfield  {author} {\bibinfo {author} {\bibfnamefont {S.}~\bibnamefont
  {Todo}}, \bibinfo {author} {\bibfnamefont {M.}~\bibnamefont {Matsumoto}},
  \bibinfo {author} {\bibfnamefont {C.}~\bibnamefont {Yasuda}}, \ and\ \bibinfo
  {author} {\bibfnamefont {H.}~\bibnamefont {Takayama}},\ }\href@noop {}
  {\bibfield  {journal} {\bibinfo  {journal} {Phys. Rev. B}\ }\textbf {\bibinfo
  {volume} {64}},\ \bibinfo {pages} {224412} (\bibinfo {year}
  {2001})}\BibitemShut {NoStop}%
\bibitem [{\citenamefont {Nakamura}\ and\ \citenamefont
  {Todo}(2002)}]{NakamuraT2002a}%
  \BibitemOpen
  \bibfield  {author} {\bibinfo {author} {\bibfnamefont {M.}~\bibnamefont
  {Nakamura}}\ and\ \bibinfo {author} {\bibfnamefont {S.}~\bibnamefont
  {Todo}},\ }\href@noop {} {\bibfield  {journal} {\bibinfo  {journal} {Phys.
  Rev. Lett.}\ }\textbf {\bibinfo {volume} {89}},\ \bibinfo {pages} {077204}
  (\bibinfo {year} {2002})}\BibitemShut {NoStop}%
\bibitem [{\citenamefont {Hatsugai}(2006)}]{Hatsugai2006}%
  \BibitemOpen
  \bibfield  {author} {\bibinfo {author} {\bibfnamefont {Y.}~\bibnamefont
  {Hatsugai}},\ }\href@noop {} {\bibfield  {journal} {\bibinfo  {journal} {J.
  Phys. Soc. Jpn.}\ }\textbf {\bibinfo {volume} {75}},\ \bibinfo {pages}
  {123601} (\bibinfo {year} {2006})}\BibitemShut {NoStop}%
\bibitem [{\citenamefont {Gu}\ and\ \citenamefont {Wen}(2009)}]{GuW2009}%
  \BibitemOpen
  \bibfield  {author} {\bibinfo {author} {\bibfnamefont {Z.-C.}\ \bibnamefont
  {Gu}}\ and\ \bibinfo {author} {\bibfnamefont {X.-G.}\ \bibnamefont {Wen}},\
  }\href@noop {} {\bibfield  {journal} {\bibinfo  {journal} {Phys. Rev. B}\
  }\textbf {\bibinfo {volume} {80}},\ \bibinfo {pages} {155131} (\bibinfo
  {year} {2009})}\BibitemShut {NoStop}%
\bibitem [{\citenamefont {Pollmann}\ \emph {et~al.}(2010)\citenamefont
  {Pollmann}, \citenamefont {Turner}, \citenamefont {Berg},\ and\ \citenamefont
  {Oshikawa}}]{PollmannTBO2010}%
  \BibitemOpen
  \bibfield  {author} {\bibinfo {author} {\bibfnamefont {F.}~\bibnamefont
  {Pollmann}}, \bibinfo {author} {\bibfnamefont {A.~M.}\ \bibnamefont
  {Turner}}, \bibinfo {author} {\bibfnamefont {E.}~\bibnamefont {Berg}}, \ and\
  \bibinfo {author} {\bibfnamefont {M.}~\bibnamefont {Oshikawa}},\ }\href@noop
  {} {\bibfield  {journal} {\bibinfo  {journal} {Phys. Rev. B}\ }\textbf
  {\bibinfo {volume} {81}},\ \bibinfo {pages} {064439} (\bibinfo {year}
  {2010})}\BibitemShut {NoStop}%
\bibitem [{\citenamefont {Li}\ and\ \citenamefont {Haldane}(2008)}]{LiH2008}%
  \BibitemOpen
  \bibfield  {author} {\bibinfo {author} {\bibfnamefont {H.}~\bibnamefont
  {Li}}\ and\ \bibinfo {author} {\bibfnamefont {F.~D.~M.}\ \bibnamefont
  {Haldane}},\ }\href@noop {} {\bibfield  {journal} {\bibinfo  {journal} {Phys.
  Rev. Lett.}\ }\textbf {\bibinfo {volume} {101}},\ \bibinfo {pages} {010504}
  (\bibinfo {year} {2008})}\BibitemShut {NoStop}%
\bibitem [{\citenamefont {Chen}\ \emph
  {et~al.}(2011{\natexlab{a}})\citenamefont {Chen}, \citenamefont {Gu},\ and\
  \citenamefont {Wen}}]{ChenGW2011}%
  \BibitemOpen
  \bibfield  {author} {\bibinfo {author} {\bibfnamefont {X.}~\bibnamefont
  {Chen}}, \bibinfo {author} {\bibfnamefont {Z.-C.}\ \bibnamefont {Gu}}, \ and\
  \bibinfo {author} {\bibfnamefont {X.-G.}\ \bibnamefont {Wen}},\ }\href@noop
  {} {\bibfield  {journal} {\bibinfo  {journal} {Phys. Rev. B}\ }\textbf
  {\bibinfo {volume} {83}},\ \bibinfo {pages} {035107} (\bibinfo {year}
  {2011}{\natexlab{a}})}\BibitemShut {NoStop}%
\bibitem [{\citenamefont {Chen}\ \emph
  {et~al.}(2011{\natexlab{b}})\citenamefont {Chen}, \citenamefont {Gu},\ and\
  \citenamefont {Wen}}]{ChenGW2011a}%
  \BibitemOpen
  \bibfield  {author} {\bibinfo {author} {\bibfnamefont {X.}~\bibnamefont
  {Chen}}, \bibinfo {author} {\bibfnamefont {Z.-C.}\ \bibnamefont {Gu}}, \ and\
  \bibinfo {author} {\bibfnamefont {X.-G.}\ \bibnamefont {Wen}},\ }\href@noop
  {} {\bibfield  {journal} {\bibinfo  {journal} {Phys. Rev. B}\ }\textbf
  {\bibinfo {volume} {84}},\ \bibinfo {pages} {235128} (\bibinfo {year}
  {2011}{\natexlab{b}})}\BibitemShut {NoStop}%
\bibitem [{\citenamefont {Duivenvoorden}\ and\ \citenamefont
  {Quella}(2013)}]{DuivenvoordenQ2013}%
  \BibitemOpen
  \bibfield  {author} {\bibinfo {author} {\bibfnamefont {K.}~\bibnamefont
  {Duivenvoorden}}\ and\ \bibinfo {author} {\bibfnamefont {T.}~\bibnamefont
  {Quella}},\ }\href@noop {} {\bibfield  {journal} {\bibinfo  {journal} {Phys.
  Rev. B}\ }\textbf {\bibinfo {volume} {87}},\ \bibinfo {pages} {125145}
  (\bibinfo {year} {2013})}\BibitemShut {NoStop}%
\bibitem [{\citenamefont {Morimoto}\ \emph {et~al.}(2014)\citenamefont
  {Morimoto}, \citenamefont {Ueda}, \citenamefont {Momoi},\ and\ \citenamefont
  {Furusaki}}]{MorimotoUMF2014}%
  \BibitemOpen
  \bibfield  {author} {\bibinfo {author} {\bibfnamefont {T.}~\bibnamefont
  {Morimoto}}, \bibinfo {author} {\bibfnamefont {H.}~\bibnamefont {Ueda}},
  \bibinfo {author} {\bibfnamefont {T.}~\bibnamefont {Momoi}}, \ and\ \bibinfo
  {author} {\bibfnamefont {A.}~\bibnamefont {Furusaki}},\ }\href@noop {}
  {\bibfield  {journal} {\bibinfo  {journal} {Phys. Rev. B}\ }\textbf {\bibinfo
  {volume} {90}},\ \bibinfo {pages} {235111} (\bibinfo {year}
  {2014})}\BibitemShut {NoStop}%
\bibitem [{\citenamefont {Duivenvoorden}\ and\ \citenamefont
  {Quella}(2012)}]{DuivenvoordenQ2012}%
  \BibitemOpen
  \bibfield  {author} {\bibinfo {author} {\bibfnamefont {K.}~\bibnamefont
  {Duivenvoorden}}\ and\ \bibinfo {author} {\bibfnamefont {T.}~\bibnamefont
  {Quella}},\ }\href@noop {} {\bibfield  {journal} {\bibinfo  {journal} {Phys.
  Rev. B}\ }\textbf {\bibinfo {volume} {86}},\ \bibinfo {pages} {235142}
  (\bibinfo {year} {2012})}\BibitemShut {NoStop}%
\bibitem [{\citenamefont {Geraedts}\ and\ \citenamefont
  {Motrunich}()}]{GeraedtsM2014arXiv}%
  \BibitemOpen
  \bibfield  {author} {\bibinfo {author} {\bibfnamefont {S.~D.}\ \bibnamefont
  {Geraedts}}\ and\ \bibinfo {author} {\bibfnamefont {O.~I.}\ \bibnamefont
  {Motrunich}},\ }\href@noop {} {\ }\Eprint
  {http://arxiv.org/abs/arXiv:1410.1580v1} {arXiv:1410.1580v1} \BibitemShut
  {NoStop}%
\bibitem [{\citenamefont {Pachos}\ and\ \citenamefont
  {Plenio}(2004)}]{PachosP2004}%
  \BibitemOpen
  \bibfield  {author} {\bibinfo {author} {\bibfnamefont {J.~K.}\ \bibnamefont
  {Pachos}}\ and\ \bibinfo {author} {\bibfnamefont {M.~B.}\ \bibnamefont
  {Plenio}},\ }\href@noop {} {\bibfield  {journal} {\bibinfo  {journal} {Phys.
  Rev. Lett.}\ }\textbf {\bibinfo {volume} {93}},\ \bibinfo {pages} {056402}
  (\bibinfo {year} {2004})}\BibitemShut {NoStop}%
\bibitem [{\citenamefont {Gorshkov}\ \emph {et~al.}(2010)\citenamefont
  {Gorshkov}, \citenamefont {Hermele}, \citenamefont {Gurarie}, \citenamefont
  {Xu}, \citenamefont {Julienne}, \citenamefont {Ye}, \citenamefont {Zoller},
  \citenamefont {Demler}, \citenamefont {Lukin},\ and\ \citenamefont
  {Rey}}]{GorshkovHGXJYZDLR2010}%
  \BibitemOpen
  \bibfield  {author} {\bibinfo {author} {\bibfnamefont {A.~V.}\ \bibnamefont
  {Gorshkov}}, \bibinfo {author} {\bibfnamefont {M.}~\bibnamefont {Hermele}},
  \bibinfo {author} {\bibfnamefont {V.}~\bibnamefont {Gurarie}}, \bibinfo
  {author} {\bibfnamefont {C.}~\bibnamefont {Xu}}, \bibinfo {author}
  {\bibfnamefont {P.~S.}\ \bibnamefont {Julienne}}, \bibinfo {author}
  {\bibfnamefont {J.}~\bibnamefont {Ye}}, \bibinfo {author} {\bibfnamefont
  {P.}~\bibnamefont {Zoller}}, \bibinfo {author} {\bibfnamefont
  {E.}~\bibnamefont {Demler}}, \bibinfo {author} {\bibfnamefont {M.~D.}\
  \bibnamefont {Lukin}}, \ and\ \bibinfo {author} {\bibfnamefont {A.~M.}\
  \bibnamefont {Rey}},\ }\href@noop {} {\bibfield  {journal} {\bibinfo
  {journal} {Nat. Phys.}\ }\textbf {\bibinfo {volume} {6}},\ \bibinfo {pages}
  {289} (\bibinfo {year} {2010})}\BibitemShut {NoStop}%
\bibitem [{\citenamefont {Nonne}\ \emph {et~al.}(2013)\citenamefont {Nonne},
  \citenamefont {Moliner}, \citenamefont {Capponi}, \citenamefont
  {Lecheminant},\ and\ \citenamefont {Totsuka}}]{NonneMCLT2013}%
  \BibitemOpen
  \bibfield  {author} {\bibinfo {author} {\bibfnamefont {H.}~\bibnamefont
  {Nonne}}, \bibinfo {author} {\bibfnamefont {M.}~\bibnamefont {Moliner}},
  \bibinfo {author} {\bibfnamefont {S.}~\bibnamefont {Capponi}}, \bibinfo
  {author} {\bibfnamefont {P.}~\bibnamefont {Lecheminant}}, \ and\ \bibinfo
  {author} {\bibfnamefont {K.}~\bibnamefont {Totsuka}},\ }\href@noop {}
  {\bibfield  {journal} {\bibinfo  {journal} {Europhys. Lett.}\ }\textbf
  {\bibinfo {volume} {102}},\ \bibinfo {pages} {37008} (\bibinfo {year}
  {2013})}\BibitemShut {NoStop}%
\bibitem [{\citenamefont {Hatsugai}\ and\ \citenamefont
  {Maruyama}(2011)}]{HatsugaiM2011}%
  \BibitemOpen
  \bibfield  {author} {\bibinfo {author} {\bibfnamefont {Y.}~\bibnamefont
  {Hatsugai}}\ and\ \bibinfo {author} {\bibfnamefont {I.}~\bibnamefont
  {Maruyama}},\ }\href@noop {} {\bibfield  {journal} {\bibinfo  {journal}
  {Europhys. Lett.}\ }\textbf {\bibinfo {volume} {95}},\ \bibinfo {pages}
  {20003} (\bibinfo {year} {2011})}\BibitemShut {NoStop}%
\bibitem [{\citenamefont {Motoyama}\ and\ \citenamefont
  {Todo}(2013)}]{MotoyamaT2013}%
  \BibitemOpen
  \bibfield  {author} {\bibinfo {author} {\bibfnamefont {Y.}~\bibnamefont
  {Motoyama}}\ and\ \bibinfo {author} {\bibfnamefont {S.}~\bibnamefont
  {Todo}},\ }\href@noop {} {\bibfield  {journal} {\bibinfo  {journal} {Phys.
  Rev. E}\ }\textbf {\bibinfo {volume} {87}},\ \bibinfo {pages} {021301(R)}
  (\bibinfo {year} {2013})}\BibitemShut {NoStop}%
\bibitem [{\citenamefont {Affleck}(1985)}]{Affleck1985a}%
  \BibitemOpen
  \bibfield  {author} {\bibinfo {author} {\bibfnamefont {I.}~\bibnamefont
  {Affleck}},\ }\href@noop {} {\bibfield  {journal} {\bibinfo  {journal} {Phys.
  Rev. Lett.}\ }\textbf {\bibinfo {volume} {54}},\ \bibinfo {pages} {966}
  (\bibinfo {year} {1985})}\BibitemShut {NoStop}%
\bibitem [{\citenamefont {Read}\ and\ \citenamefont
  {Sachdev}(1990)}]{ReadS1990}%
  \BibitemOpen
  \bibfield  {author} {\bibinfo {author} {\bibfnamefont {N.}~\bibnamefont
  {Read}}\ and\ \bibinfo {author} {\bibfnamefont {S.}~\bibnamefont {Sachdev}},\
  }\href@noop {} {\bibfield  {journal} {\bibinfo  {journal} {Phys. Rev. B}\
  }\textbf {\bibinfo {volume} {42}},\ \bibinfo {pages} {4568} (\bibinfo {year}
  {1990})}\BibitemShut {NoStop}%
\bibitem [{\citenamefont {Harada}\ \emph {et~al.}(2003)\citenamefont {Harada},
  \citenamefont {Kawashima},\ and\ \citenamefont {Troyer}}]{HaradaKT2003}%
  \BibitemOpen
  \bibfield  {author} {\bibinfo {author} {\bibfnamefont {K.}~\bibnamefont
  {Harada}}, \bibinfo {author} {\bibfnamefont {N.}~\bibnamefont {Kawashima}}, \
  and\ \bibinfo {author} {\bibfnamefont {M.}~\bibnamefont {Troyer}},\
  }\href@noop {} {\bibfield  {journal} {\bibinfo  {journal} {Phys. Rev. Lett.}\
  }\textbf {\bibinfo {volume} {90}},\ \bibinfo {pages} {117203} (\bibinfo
  {year} {2003})}\BibitemShut {NoStop}%
\bibitem [{Note1()}]{Note1}%
  \BibitemOpen
  \bibinfo {note} {The differentiation erases a summation as $\partial _{\phi
  _\alpha }\DOTSB \sum@ \slimits@ _\beta e^{-i\phi _\beta } = -ie^{-i\phi
  _\alpha },$ and produces the denominator, $N$. The numerator, $M$, comes from
  the Leibniz's product rule.}\BibitemShut {Stop}%
\bibitem [{\citenamefont {Todo}(2013)}]{Todo2013a}%
  \BibitemOpen
  \bibfield  {author} {\bibinfo {author} {\bibfnamefont {S.}~\bibnamefont
  {Todo}},\ }in\ \href@noop {} {\emph {\bibinfo {booktitle} {Strongly
  Correlated Systems: Numerical Methods (Springer Series in Solid-State
  Sciences)}}},\ \bibinfo {editor} {edited by\ \bibinfo {editor} {\bibfnamefont
  {A.}~\bibnamefont {Avella}}\ and\ \bibinfo {editor} {\bibfnamefont
  {F.}~\bibnamefont {Mancini}}}\ (\bibinfo  {publisher} {Springer-Verlag},\
  \bibinfo {address} {Berlin},\ \bibinfo {year} {2013})\ pp.\ \bibinfo {pages}
  {153--184}\BibitemShut {NoStop}%
\bibitem [{\citenamefont {Beard}\ and\ \citenamefont
  {Wiese}(1996)}]{BeardW1996}%
  \BibitemOpen
  \bibfield  {author} {\bibinfo {author} {\bibfnamefont {B.~B.}\ \bibnamefont
  {Beard}}\ and\ \bibinfo {author} {\bibfnamefont {U.~J.}\ \bibnamefont
  {Wiese}},\ }\href@noop {} {\bibfield  {journal} {\bibinfo  {journal} {Phys.
  Rev. Lett.}\ }\textbf {\bibinfo {volume} {77}},\ \bibinfo {pages} {5130}
  (\bibinfo {year} {1996})}\BibitemShut {NoStop}%
\bibitem [{\citenamefont {Todo}\ and\ \citenamefont {Kato}(2001)}]{TodoK2001}%
  \BibitemOpen
  \bibfield  {author} {\bibinfo {author} {\bibfnamefont {S.}~\bibnamefont
  {Todo}}\ and\ \bibinfo {author} {\bibfnamefont {K.}~\bibnamefont {Kato}},\
  }\href@noop {} {\bibfield  {journal} {\bibinfo  {journal} {Phys. Rev. Lett.}\
  }\textbf {\bibinfo {volume} {87}},\ \bibinfo {pages} {047203} (\bibinfo
  {year} {2001})}\BibitemShut {NoStop}%
\bibitem [{\citenamefont {Quan}\ \emph {et~al.}(2006)\citenamefont {Quan},
  \citenamefont {Song}, \citenamefont {Liu}, \citenamefont {Zanardi},\ and\
  \citenamefont {Sun}}]{QuanSLZS2006}%
  \BibitemOpen
  \bibfield  {author} {\bibinfo {author} {\bibfnamefont {H.~T.}\ \bibnamefont
  {Quan}}, \bibinfo {author} {\bibfnamefont {Z.}~\bibnamefont {Song}}, \bibinfo
  {author} {\bibfnamefont {X.~F.}\ \bibnamefont {Liu}}, \bibinfo {author}
  {\bibfnamefont {P.}~\bibnamefont {Zanardi}}, \ and\ \bibinfo {author}
  {\bibfnamefont {C.~P.}\ \bibnamefont {Sun}},\ }\href@noop {} {\bibfield
  {journal} {\bibinfo  {journal} {Phys. Rev. Lett.}\ }\textbf {\bibinfo
  {volume} {96}},\ \bibinfo {pages} {140604} (\bibinfo {year}
  {2006})}\BibitemShut {NoStop}%
\bibitem [{\citenamefont {Schwandt}\ \emph {et~al.}(2009)\citenamefont
  {Schwandt}, \citenamefont {Alet},\ and\ \citenamefont
  {Capponi}}]{SchwandtAC2009}%
  \BibitemOpen
  \bibfield  {author} {\bibinfo {author} {\bibfnamefont {D.}~\bibnamefont
  {Schwandt}}, \bibinfo {author} {\bibfnamefont {F.}~\bibnamefont {Alet}}, \
  and\ \bibinfo {author} {\bibfnamefont {S.}~\bibnamefont {Capponi}},\
  }\href@noop {} {\bibfield  {journal} {\bibinfo  {journal} {Phys. Rev. Lett.}\
  }\textbf {\bibinfo {volume} {103}},\ \bibinfo {pages} {170501} (\bibinfo
  {year} {2009})}\BibitemShut {NoStop}%
\bibitem [{\citenamefont {Kolodrubetz}(2014)}]{Kolodrubetz2014}%
  \BibitemOpen
  \bibfield  {author} {\bibinfo {author} {\bibfnamefont {M.}~\bibnamefont
  {Kolodrubetz}},\ }\href@noop {} {\bibfield  {journal} {\bibinfo  {journal}
  {Phys. Rev. B}\ }\textbf {\bibinfo {volume} {89}},\ \bibinfo {pages} {045107}
  (\bibinfo {year} {2014})}\BibitemShut {NoStop}%
\bibitem [{\citenamefont {Chandrasekharan}\ and\ \citenamefont
  {Wiese}(1999)}]{ChandrasekharanW1999}%
  \BibitemOpen
  \bibfield  {author} {\bibinfo {author} {\bibfnamefont {S.}~\bibnamefont
  {Chandrasekharan}}\ and\ \bibinfo {author} {\bibfnamefont {U.-J.}\
  \bibnamefont {Wiese}},\ }\href@noop {} {\bibfield  {journal} {\bibinfo
  {journal} {Phys. Rev. Lett.}\ }\textbf {\bibinfo {volume} {83}},\ \bibinfo
  {pages} {3116} (\bibinfo {year} {1999})}\BibitemShut {NoStop}%
\bibitem [{\citenamefont {Bauer}\ \emph {et~al.}(2011)\citenamefont {Bauer},
  \citenamefont {Carr}, \citenamefont {Evertz}, \citenamefont {Feiguin},
  \citenamefont {Freire}, \citenamefont {Fuchs}, \citenamefont {Gamper},
  \citenamefont {Gukelberger}, \citenamefont {Gull}, \citenamefont {Guertler},
  \citenamefont {Hehn}, \citenamefont {Igarashi}, \citenamefont {Isakov},
  \citenamefont {Koop}, \citenamefont {Ma}, \citenamefont {Mates},
  \citenamefont {Matsuo}, \citenamefont {Parcollet}, \citenamefont {Pawlowski},
  \citenamefont {Picon}, \citenamefont {Pollet}, \citenamefont {Santos},
  \citenamefont {Scarola}, \citenamefont {Schollw\"ock}, \citenamefont {Silva},
  \citenamefont {Surer}, \citenamefont {Todo}, \citenamefont {Trebst},
  \citenamefont {Troyer}, \citenamefont {Wall}, \citenamefont {Werner},\ and\
  \citenamefont {Wessel}}]{ALPS2011}%
  \BibitemOpen
  \bibfield  {author} {\bibinfo {author} {\bibfnamefont {B.}~\bibnamefont
  {Bauer}}, \bibinfo {author} {\bibfnamefont {L.~D.}\ \bibnamefont {Carr}},
  \bibinfo {author} {\bibfnamefont {H.~G.}\ \bibnamefont {Evertz}}, \bibinfo
  {author} {\bibfnamefont {A.}~\bibnamefont {Feiguin}}, \bibinfo {author}
  {\bibfnamefont {J.}~\bibnamefont {Freire}}, \bibinfo {author} {\bibfnamefont
  {S.}~\bibnamefont {Fuchs}}, \bibinfo {author} {\bibfnamefont
  {L.}~\bibnamefont {Gamper}}, \bibinfo {author} {\bibfnamefont
  {J.}~\bibnamefont {Gukelberger}}, \bibinfo {author} {\bibfnamefont
  {E.}~\bibnamefont {Gull}}, \bibinfo {author} {\bibfnamefont {S.}~\bibnamefont
  {Guertler}}, \bibinfo {author} {\bibfnamefont {A.}~\bibnamefont {Hehn}},
  \bibinfo {author} {\bibfnamefont {R.}~\bibnamefont {Igarashi}}, \bibinfo
  {author} {\bibfnamefont {S.~V.}\ \bibnamefont {Isakov}}, \bibinfo {author}
  {\bibfnamefont {D.}~\bibnamefont {Koop}}, \bibinfo {author} {\bibfnamefont
  {P.~N.}\ \bibnamefont {Ma}}, \bibinfo {author} {\bibfnamefont
  {P.}~\bibnamefont {Mates}}, \bibinfo {author} {\bibfnamefont
  {H.}~\bibnamefont {Matsuo}}, \bibinfo {author} {\bibfnamefont
  {O.}~\bibnamefont {Parcollet}}, \bibinfo {author} {\bibfnamefont
  {G.}~\bibnamefont {Pawlowski}}, \bibinfo {author} {\bibfnamefont {J.~D.}\
  \bibnamefont {Picon}}, \bibinfo {author} {\bibfnamefont {L.}~\bibnamefont
  {Pollet}}, \bibinfo {author} {\bibfnamefont {E.}~\bibnamefont {Santos}},
  \bibinfo {author} {\bibfnamefont {V.~W.}\ \bibnamefont {Scarola}}, \bibinfo
  {author} {\bibfnamefont {U.}~\bibnamefont {Schollw\"ock}}, \bibinfo {author}
  {\bibfnamefont {C.}~\bibnamefont {Silva}}, \bibinfo {author} {\bibfnamefont
  {B.}~\bibnamefont {Surer}}, \bibinfo {author} {\bibfnamefont
  {S.}~\bibnamefont {Todo}}, \bibinfo {author} {\bibfnamefont {S.}~\bibnamefont
  {Trebst}}, \bibinfo {author} {\bibfnamefont {M.}~\bibnamefont {Troyer}},
  \bibinfo {author} {\bibfnamefont {M.~L.}\ \bibnamefont {Wall}}, \bibinfo
  {author} {\bibfnamefont {P.}~\bibnamefont {Werner}}, \ and\ \bibinfo {author}
  {\bibfnamefont {S.}~\bibnamefont {Wessel}},\ }\href@noop {} {\bibfield
  {journal} {\bibinfo  {journal} {J. Stat. Mech.: Theo. Exp.}\ ,\ \bibinfo
  {pages} {P05001}} (\bibinfo {year} {2011})}\BibitemShut {NoStop}%
\bibitem [{ALP()}]{ALPSweb}%
  \BibitemOpen
  \href@noop {} {\enquote {\bibinfo {title} {{http://alps.comp-phys.org/}},}\
  }\BibitemShut {NoStop}%
\end{thebibliography}%
\end{document}